\documentclass[preprint]{aastex}
\everymath{\rm}
\everydisplay{\sf}
\received{}
\revised{}
\accepted{}
\cpright{}{}\journalid{}{}
\articleid{}{}
\paperid{}
\slugcomment{To appear in {\it The Astrophysical Journal}}
\shortauthors{Kwitter \& Henry}
\shorttitle{S, Cl, \& Ar in PNe}
\begin{document}
\title{Sulfur, Chlorine, \& Argon Abundances in Planetary Nebulae. I: Observations and Abundances in a
Northern Sample}

\author{K.B. Kwitter\footnote{Visiting Astronomer, Kitt Peak
National Observatory, National Optical Astronomy Observatories,
which is operated by the Association of Universities for Research in
Astronomy, Inc. (AURA) under cooperative agreement with the
National Science Foundation.}}

\affil{Department of Astronomy, Williams College, Williamstown, MA
01267; kkwitter@williams.edu}

\and

\author {R.B.C. Henry$^1$}

\affil{Department of Physics \& Astronomy, University of Oklahoma,
Norman, OK 73019; henry@mail.nhn.ou.edu}

\begin{abstract}

This paper is the first of a series specifically studying the
abundances of sulfur, chlorine, and argon in Type~II planetary nebulae
(PNe) in the Galactic disk. Ratios of S/O, Cl/O, and Ar/O constitute
important tests of differential nucleosynthesis of these elements and
serve as strict constraints on massive star yield predictions. We
present new ground-based optical spectra extending from 3600-9600{\AA}
for a sample of 19 Type~II northern PNe. This range includes the
strong near infrared lines of [S~III] $\lambda\lambda$9069,9532, which
allows us to test extensively their effectiveness as sulfur abundance
indicators. We also introduce a new, model-tested ionization correction
factor for sulfur. For the present sample, we find average values of
S/O=1.2E-2$\pm$0.71E-2, Cl/O=3.3E-4$\pm$1.6E-4, and
Ar/O=5.0E-3$\pm$1.9E-3.

\end{abstract}


\clearpage

\section{Introduction}

As a galaxy evolves chemically, interstellar gas is cycled through
stars and becomes enriched with metals as these stars convert hydrogen
and helium into heavier elements via nuclear processing. Much of our
understanding of this whole scheme comes about through the interplay
between model-predicted and observed heavy element abundance ratios as
a function of some gauge (e.g., metallicity) of the extent to which a
system has evolved.

The chemical evolution models that are used to interpret observed
abundance data rely sensitively on predicted stellar production rates,
or yields, of individual heavy elements, and these rates are in turn
inferred from detailed stellar evolution models. Current examples of
stellar models with yield predictions include those by Woosley \&
Weaver (1995), Maeder (1992), and Nomoto et al. (1997a) for massive
stars, and by Marigo et al. (1998) and van den Hoek \& Groenewegen
(1997) for intermediate mass stars. Each of these teams uses
state-of-the-art stellar physics, along with poorly constrained
assumptions about mixing length, convective overshooting, and mass
loss, to predict the new amounts of numerous isotopes that are added
to the interstellar medium through winds, supernovae, and planetary
nebulae as stars evolve and perish.

Model predictions of heavy element abundance ratios as functions of
metallicity are subsequently compared with observations of these same
ratios in either the interstellar medium or in stars. Assessing the
success of different theoretical stellar yield predictions with
observations sharpens our understanding of production rates of heavy
elements. Interstellar medium abundances from, e.g., H~II regions,
testify to the current chemical composition at particular locations in
the Galaxy. Stellar abundances archive the composition of the places
and times of their formation. Planetary nebulae (PNe) are especially
useful because they can be used to infer both current and past
compositions: some of their element abundances can be altered by
nucleosynthesis in the progenitor; other elements, impossible to make
or destroy in the conditions within the progenitor core or envelope,
should remain at original stellar birth levels. Thus, abundance
studies of PNe originating from stars representing different epochs
can reveal an evolutionary picture of how element ratios have changed
over time.

One important method for testing nucleosynthetic theory is to look at
the evolutionary change of one heavy element with respect to
another. In interstellar abundance studies, O/H is almost always taken
as the gauge of metallicity, and so abundance ratios such as S/O
become measures not of absolute metallicity changes but of
differential changes between two elements as chemical evolution
progresses.

The use of PNe to measure the original stellar abundances of elements
such as sulfur is possible because PN progenitors, intermediate-mass
stars with masses between 1-8~M$_{\odot}$, lack sufficient mass to
synthesize them, and thus the levels found in the material shed by the
dying star to form the nebula are a measure of what was present in the
interstellar material when the star itself formed.

This paper is the first in a series of five that strives to test
differential nucleosynthesis predictions for oxygen, sulfur, chlorine,
and argon by measuring the abundances of these elements in a large
sample ($>$60 objects) of PNe residing in the Galactic disk and that
together represent a significant range in galactocentric distance. The
latter point means that, because of the metallicity gradient in the
disk, we can explore objects over a modest range in metallicity as
measured by O/H, and discover whether the ratios of S/O, Cl/O, and
Ar/O behave according to predictions. Various aspects of the
abundances of these elements have been explored by other
authors including Aller \& Czyzak (1983), Aller \& Keyes (1987),
Barker(1978a,b), Dennefeld \& Stasi{\'n}ska (1983), Dinerstein (1980),
Freitas-Pacheco (1993), Freitas-Pacheco, Maciel, \& Costa (1992),
Kingsburgh \& Barlow (1994), K{\"o}ppen, Acker, \& Stenholm (1991),
Maciel \& K{\"o}ppen (1994), Maciel \& Chiappini (1994), Maciel \& Quireza (1999), and Peimbert \&
Torres-Peimbert (1987).

Our survey includes the important near-infrared (NIR) emission lines
of [S~III] $\lambda\lambda$9069,9532. Pioneering work on observing
[S~III] in PNe was carried out by Barker (1978b, 1980b, 1983) and by
Dinerstein (1980), both of whom used a photoelectric scanner to
measure $\lambda$9532 at $\sim$20{\AA } resolution. We are able to take
advantage of the improvement in detector technology to achieve spatial
resolution that measures precisely the same region in the nebula in
the NIR as in the optical, as well as spectral range that
enables observation of both $\lambda$9069 and $\lambda$9532 (see
section 3), and spectral resolution that allows separation of $\lambda$9532 
from Paschen 8 at $\lambda$9546.

Here we present new spectrophotometric measurements of 19 northern
hemisphere Type~II PNe between 3600-9600{\AA}. Type~II PNe were chosen
because there is substantial evidence through their kinematics that
these objects are part of a young population orbiting very near the
plane of the Galactic disk (Peimbert 1990). At the same time, their
progenitor stars are considered to be of insufficient mass to allow
oxygen depletion or neon enrichment (through ON cycling or carbon
burning, respectively) during the star's life. Hence they provide
current measurements of interstellar abundances in the disk.

Using 5-level atom routines along with model-tested ionization
correction factors for certain elements, we calculate and report the
nebular abundances of He, N, O, Ne, S, Cl, and Ar in 18 objects. (For
one of the 19 observed PNe, NGC 6894, we did not detect measureable
[O~III] $\lambda$4363, and so we could not perform a complete
abundance analysis.) We focus in particular on the resulting abundance
ratios of S/O, Cl/O, and Ar/O. Combining $\lambda\lambda$9069,9532
with the $\lambda$6312 auroral line of [S~III], we are able to derive
a [S~III] temperature to measure abundances of S$^{+2}$. Our hope is
to verify or improve upon the S abundances heretofore determined using
only the more accessible but weaker [S~III] $\lambda$6312 line to
derive S$^{+2}$. We also introduce a new model-tested ionization
correction factor for sulfur, derived in light of new
atomic data and sensitive to matter-boundedness.

Papers~IIa and IIb in the series are a continuation of our study,
using a sample of 48 southern hemisphere Type~II PNe; the former will
present the observations, the latter will describe the abundance
analysis. Paper~III will present abundance results for Type~II PNe
with available ISO mid-IR fine-structure lines of S$^{+3}$. Paper~IV
will present new S, Ar, and Cl abundances for a set of Type~II PNe
whose other abundances were included in a previous series of
papers. Finally, Paper~V will make use of all of our amassed data plus
information on planetary nebula distances to interpret our derived
abundance ratios in light of published stellar yields and our own
chemical evolution models. One principal goal of our study is to look
for the signature of contributions from Type~Ia supernovae, which,
according to models by Nomoto et al. (1997b), generate a significant
amount of S, Cl, and Ar per event. While the rate of Type~Ia events is
thought to be several times below that of Type~II events (Cappellaro, Evans, \& Turatto 1999), we would
like to ascertain whether the former objects do, in fact, play an
important role in the chemical evolution of S, Cl, and Ar.

The organization of this paper is as follows. Section~2 contains a
discussion of the spectral data, including a sample plot with
important line identifications, and our reported line
strengths. Section~3 discusses our methods for deriving the ionic and
elemental abundances and includes comparisons and plots of our
results. Section~4 is a summary of our findings for this sample of
objects.

\section{Observations and Reductions}

\subsection{Observations}

Observations were obtained at KPNO during 1999 28 June - 1 July using
the Goldcam CCD spectrometer attached to the 2.1m telescope. The chip
was a Ford 3K $\times$ 1K CCD with 15$\mu$ pixels. We used a slit that
was 5$\arcsec$ wide and extended 285$\arcsec$ in the E-W direction,
with a spatial scale of 0$\farcs$78/pixel. With a combination of two
gratings, we obtained spectral coverage from 3700-9600\AA\ with
overlapping coverage from $\sim$5750 - 6750\AA.  Using grating $\#$240
with a WG 345 order-separation filter, wavelength dispersion was 1.5 \AA/pixel
($\sim$8 \AA\ FWHM resolution) for the blue. For the red we used
grating $\#$58 with an OG530 order-separation filter, yielding 1.9 \AA/pixel
($\sim$10 \AA\ FWHM resolution).  Table~1 lists the objects observed,
their angular sizes, and the exposure times in seconds for the blue
and red grating configurations.  Most of these PNe are relatively
small in angular size; therefore, we placed the Goldcam slit on the
brightest part of the nebula as seen on the acquisition screen,
avoiding the central star if it was visible. We obtained the usual
bias and twilight flat-field frames each night, along with HeNeAr
comparison spectra for wavelength calibration and standard star
spectra for sensitivity calibration.

The thinned red chip produces interference fringes visible in the red.
In our red spectra the fringes appear at the $\pm$1\% level at
$\sim$7500\AA\ and increase in amplitude with increasing wavelength:
$\pm$1.5\% at 8000\AA, $\pm$4.5\% at 8500\AA, $\pm$6\% at
9000\AA. Even at their worst, {\it i.e.\/}, at $\sim$$\lambda$9500,
the longest wavelength we measure, the fringe amplitude reaches only
about $\pm$7\%, and we note this additional
uncertainty in our line intensities longward of $\sim$7500\AA.

A typical spectrum is shown in Fig.~1, combined here into a single
plot. In this nebula, M1-57, lines from multiple ions of S, Ar, and Cl
are identified, along with other important spectral features.

The original spectra were reduced in the standard fashion using
IRAF\footnote{IRAF is distributed by the National Optical Astronomy
Observatories, which is operated by the Association of Universities
for Research in Astronomy, Inc. (AURA) under cooperative agreement
with the National Science Foundation.}. We used tasks in the {\it
kpnoslit\/} package to convert these two-dimensional spectra to
one dimension by collapsing data along the slit. 

\subsection{Line Strengths}

Line strengths were measured using {\it splot} in IRAF and are
reported in Tables~2A-D.  Fluxes uncorrected for reddening are
presented in columns labeled F($\lambda$), where these flux values
have been normalized to H$\beta$=100 using our observed value of log
F$_{H\beta}$ shown in the last row of the table.  These line strengths
in turn were corrected for reddening by assuming that the relative
strength of H$\alpha$/H$\beta$=2.86 (Osterbrock 1989; Table~4.4) and computing the logarithmic
extinction quantity {\it c} shown in the penultimate line of the
table.  Values for the reddening coefficients, f($\lambda$), are
listed in column~(2), where we employed the extinction curve of Savage
\& Mathis (1979). Intensities are corrected by multiplying the
observed ratio relative to H$\beta$ by dexp[{\it c}f($\lambda$)]. To
check the validity of the values of {\it c} derived from
H$\alpha$/H$\beta$ we calculated values using the ratio of P10
$\lambda$9014 and P8 $\lambda$9546, each with respect to
H$\beta$. Table~3A contains all of the calculated values for {\it c},
which are also shown graphically in Fig.~2.  As can be seen, the
agreement is generally quite good between H$\alpha$ and Paschen extinction $c$ measurements, providing support for our spectrum reduction and merging techniques.

The columns headed I($\lambda$) in Tables~2A-D list our final, corrected line
strengths, again normalized to H$\beta$=100.  In general, intensities
of strong lines have measurement uncertainties $\le$10\%; single
colons indicate uncertainties of more than $\sim$25\%, and double colons
denote uncertainties exceeding $\sim$50\%.

Finally, as a check on the accuracy of our final line strengths, in
Table~3B we compare observed and theoretical values for a number of
line ratios which are set by atomic constants. The first column lists
the object name while the subsequent seven columns give the observed
ratios, where each ratio is defined in the table footnote. The last
two rows of the table give the observed mean and standard deviation,
and theoretical value of each ratio. Agreement is reasonable for all
but the [Ne~III] ratio, which may be affected by incomplete subtraction
of H$\epsilon$. The closeness of the other ratios to their theoretical
values confirms our general claim of line strength uncertainty of
$\pm$10\% for strong and moderately strong lines.

\section{Calculations}

\subsection{The General Scheme}

Electron temperatures and densities, ionic abundances, and total
elemental abundances were computed from our measured line strengths using the program ABUN (written by
R.B.C.H.), which features a 5-level atom routine along with ionization
correction factors (ICF) either taken from the literature or derived
below. Table~4 summarizes the ions observed, the wavelengths of the emission
lines used to obtain ionic abundances, temperatures, and densities,
and the sources of the atomic data used in ABUN. 

Ionic abundances are calculated directly from line strengths; the former are then added together and
their sum multiplied by an appropriate ICF, i.e., the ratio of
total elemental abundance to the sum of observable ions, to correct for
unseen ions. This procedure can be expressed analytically for the number
abundance of element X as follows:

\begin{equation}
N(X) =
\left\{{\sum^{obs}}{{I_{\lambda}}\over{\epsilon_{\lambda}(T_e,N_e)}}\right\}
\cdot ICF(X).
\end{equation}
  
Since we are generally interested in determining ionic abundances with
respect to H$^+$, I$_{\lambda}$ is entered as a value which has been
normalized with respect to H$\beta$, and hence an H$\beta$ generation
rate is included in the denominator of Eq.~1.

We prefer this approach to modeling each nebula individually
using a photoionization code for the following reasons. First, nearly
all PNe have irregular geometries which cannot easily be included in
such a model. Second, unless the entire nebula lies within the
spectroscopic slit, observed lines-of-sight cut through only a portion
of the nebula, and the resulting data are not likely to be equivalent
to those obtained with whole-nebula observations; photoionization
models representing the latter do not necessarily produce line
strengths which can be meaningfully compared with observed
ones. Third, applying our method to output from a range of
photoionization models representing large regions of parameter space
returns abundances that are quite consistent with the original model
input abundance set.  Finally, in previous work (Henry, Kwitter, \&
Bates 2000; Kwitter \& Henry 1996, 1998; Henry \& Kwitter 1998; Henry,
Kwitter, \& Howard 1996), we have carefully modeled individual
line-of-sight sections of nebulae and calculated a model-determined
correction factor which we then applied to abundances obtained using
the standard ICF approach. In nearly all cases these model-determined
corrections were found to be minor compared with other sources of
uncertainty.

We want to draw
particular attention to our use of the NIR lines of [S~III]
$\lambda\lambda$9069,9532 for determining S$^{+2}$ abundances. It is
the use of high-quality measurements of these lines in a large PN
sample that distinguishes our study from previous ones that have
included sulfur abundance measurements. While these nebular lines are
much stronger than the auroral line at $\lambda$6312, they ostensibly
suffer from effects of telluric absorption and emission, effects that
are extremely difficult to remove. The main problem is that the
atmospheric features are molecular in origin (primarily OH) and are
often very narrow compared with the widths of the nebular lines, since
the latter may be broadened by turbulent motions within the nebula
itself. This problem has been extensively analyzed by Stevenson
(1994), who proposed a rigorous remedy that includes
the use of relatively featureless comparison stellar spectra to
determine the location and intensity of each molecular band. However,
because such stellar spectra were not available to us, we used the
fact that the ratio of the two NIR lines of [S~III] is determined by
atomic physics, is independent of environment, and is equal to 2.48,
based on the S$^{+2}$ atomic data cited in Table~4. Thus, if the
observed intensity ratio of $\lambda$9532/$\lambda$9069 in an object
was greater than or equal to 2.48, we assumed that the $\lambda$9532
line was less affected by atmospheric absorption and we used it in the
ionic abundance calculation for S$^{+2}$. Alternatively, when the
observed line ratio was less than 2.48, we used the $\lambda$9069 line
for the ionic abundance determination, assuming it to be the less
affected.

Corrected line strengths from Table~2A-D were read into ABUN and
calculations were made as follows. Electron temperatures and densities
were calculated in the standard fashion using ratios of lines from
auroral and nebular transitions sensitive to temperature, density, or
both. Abundances of observed ions were calculated by dividing the
energy production rate per ion at the observed wavelength
$\epsilon_{\lambda}$(T$_e$,N$_e$) into the observed line intensity
I$_{\lambda}$ for each of the emission lines listed in Table~4. In
calculating ionic abundances, the [O~III] temperature was used for the
high ionization species (O$^{+2}$, Ne$^{+2}$, Cl$^{+2}$, Cl$^{+3}$,
Ar$^{+2}$, and Ar$^{+3}$), the [N~II] temperature was used for the low
ionization species (O$^o$, O$^+$, N$^+$, and S$^+$), and the [S~III]
temperature was used for S$^{+2}$. The sum of observed ions of one element was then multiplied by the appropriate ICF, as shown in eq.~1.

\subsection{Ionization Correction Factors}

ICFs for He, O, N, Ne, and Ar have been reasonably well established by
numerous investigators in the past. After testing candidates with
photoionization models, we are convinced that the ICFs for these
elements reported by Kingsburgh \& Barlow (1994) in their study of
abundances in southern PNe provide reliable results, so we have
adopted them for determining final elemental abundances below.

\subsubsection{The Ionization Correction Factors for Sulfur and Chlorine}

The ICF for sulfur is a different matter. Initial attempts to devise
such a factor were based primarily on the fact that ionization
potentials for S$^{+2}$ and O$^+$ are nearly the same, i.e. 35.1eV and
34.8eV, respectively. Hence, Peimbert \& Costero (1969) used
\begin{equation}
S/H = \frac{S^+ + S^{+2}}{H^+}\frac{O}{O^+},
\end{equation}
where each atom or ion symbol represents a number abundance. The second fraction on the right hand side is the sulfur
ionization correction factor ICF(S) when lines of both S$^+$ and
S$^{+2}$ are observed.  This form was later employed by Barker (1978b), but Barker, as well as Pagel (1978),
noticed that when this form was used, resulting S/H ratios increased
systematically with O/O$^+$, indicating the occurrence of an
overcorrection for unobserved S$^{+3}$ when the O/O$^+$ ratio is
relatively large, e.g. in highly ionized and/or matter-bounded nebulae.

Stasi{\'n}ska (1978), in a study of H~II regions, was the first to use
photoionization models to derive a correction factor, the form of
which is:
\begin{equation}
ICF(S) = [1-(O^+/O)^{\alpha}]^{-1/\alpha},
\end{equation}
where $\alpha=3$. French (1981) employed this form in a study of PNe,
but favored $\alpha=2$, a value subsequently adopted by Dennefeld \&
Stasi{\'n}ska (1983) in another H~II region study following an
analysis based upon an updated model grid by Stasi{\'n}ska (1982)
which this time included the effects of charge exchange on the
ionization structure of sulfur. Garnett (1989) also calculated a
separate photoionzation model grid for a study of H~II regions and
found a behavior for ICF(S) which fell between the curves for
$\alpha=2$ and $\alpha=3$. Finally, two papers determined the
abundance of the S$^{+3}$ ion, the principal unseen species. Natta,
Panagia, \& Preite-Martinez (1980) used photoionization and
recombination cross-sections then available to calculate the ratio of
S$^{+3}$/S$^{+2}$ by assuming ionization equilibrium and obtaining the
value of the denominator directly from observations. On the other
hand, Dinerstein (1980) observed the S$^{+3}$ ion abundance directly
by measuring [S~IV] 10.5$\mu$m in a sample of 12 PNe and found that using O/O$^+$ as the ICF(S) overpredicted the S$^{+3}$ abundance.

We felt there was a need to reexamine the ICF(S) issue primarily because of
changes in atomic data for sulfur. First, state-of-the-art
photoionization cross-sections (Verner et al. 1996), which are based
upon the results of the Opacity Project (Seaton et al. 1992) have
nearly doubled for energies slightly above threshold for the S$^+$ and
S$^{+2}$ ions since the early values computed by Chapman \& Henry
(1971) were employed by Stasi{\'n}ska (1978, 1982). Of course
radiative recombination rates which are based upon these
cross-sections have also changed. Second, charge exchange rates appear
to rival those of radiative recombination. For example, using rate
coefficients from Butler \& Dalgarno (1980) for the reaction $S^{+3} +
H^o \rightarrow S^{+2} + H^+$, radiative recombination rates from
Aldrovandi \& Pequignot (1973), and assuming that the ratio of
electron density to H$^o$ density is 10$^3$ (typical for ionized
nebulae), we find that the rates are roughly equal, and so it is
imperative that charge exchange be included in the development of
ICF(S). Both Stasi{\'n}ska (1982) and Garnett (1989) used the Butler
and Dalgarno charge exchange rates in their models, although the
effects of this process were not, of course, part of the ICF(S)
which was based simply on ionization potentials (eq.~2), and used in the early work on sulfur. Finally,
there is the problem introduced by dielectronic recombination (DR)
reactions involving complex ions such as those of sulfur. Essentially,
DR for sulfur is the wild card in the analysis (G. Ferland, private
communication), and the situation is discussed in Ali et
al. (1991). Although low-temperature rates have yet to be calculated
for ions of sulfur, the photoionization model code CLOUDY (Ferland 1996) contains estimated cross-sections and
allows rough estimates of the effects of DR to be made. We ran two
models, both for stellar effective temperature of
100,000K, total density of 1000~cm$^{-3}$, and solar abundances. One
model included DR in the calculation, while the other did not. We
found that the fraction of sulfur in the S$^+$ and S$^{+2}$ stages
increased by 34\% and 14\%, respectively, when DR was part of the
calculation, and thus the ICF(S) would be expected to be significantly lower when this process is included.

We therefore have carried out our own investigation of ICF(S) by calculating a
new grid of photoionization models using CLOUDY version 94.  We computed models with central star blackbody effective
temperatures of 35,000K, 50,000K, 75,000K, 100,000K, and 150,000K, and
for each of these temperatures we employed two density regimes,
10~cm$^{-3}$ and 1000~cm$^{-3}$. For each temperature-density combination we calculated several models which covered a range of matter-boundedness. Elemental abundances were set at solar values in all cases. Each model included up-to-date atomic data as well as effects of charge transfer and dielectronic recombination processes, most notably those relevant to ions of sulfur.

Values of S/(S$^+$+S$^{+2}$) and O/O$^+$ for each model were formed by extracting the
relative ionic abundances for S$^+$, S$^{+2}$, and O$^+$ over the entire nebula from the output of each model and combining these results with the input elemental abundances for S and O. Fig.~3 is a logarithmic plot of the first ratio, which equals the ICF(S), against the second one, where each open circle represents one model from our grid. The solid line shows a third
order fit through the points with the form $ICF(S)=dexp[-.017+(.18 \times \beta) -(.11 \times \beta^2) +(.072
\times \beta^3)]$, and $\beta=log(O/O^{+})$. The solid horizontal line toward the
top of the graph indicates the range of the observed values of
log(O/O$^+$) in our PN sample. The curves that result
from using the ICF(S) expressions given by Peimbert \& Costero (1969;
eq.~2 above; dot-dashed line), French (1981; dotted line; eq.~3 with
$\alpha=2$) and Stasi{\'n}ska (1978; dashed line; eq.~3 with
$\alpha=3$) are also shown in Fig.~3 for comparison. It would
appear that the Stasi{\'n}ska expression is similar to ours, although
at large O/O$^+$ values even that relation
will tend to overestimate ICF(S) and produce systematically larger values for
S abundances relative to our scheme. This result is probably due to the estimated effects of DR which are included in our analysis for the first time. In our
abundance calculations for S to follow, we use our analytical fit
given above as the expression for ICF(S).

Obtaining a suitable ICF for chlorine proved to be a more
straightforward task than the one just described for sulfur. By studying the ionization structure of our models we found that the observed ions of Cl$^{+2}$
and Cl$^{+3}$ roughly coexist with He$^+$, and thus
Cl/(Cl$^{+2}$+Cl$^{+3}$)=He/He$^+$=ICF(Cl) seems appropriate. We
adopt this form in the abundance calculations below.

We now summarize the ICFs we employed in our analysis:
 
\begin{mathletters}
\begin{eqnarray}
ICF(He) &=& {{He}\over{He^+}+He^{+2}} = 1.0, \\ 
ICF(O) &=&{{O}\over{O^{+2}+O^+}} = {{He^+ + He^{+2}}\over{He^+}},\\ 
ICF(N) &=&{{N}\over{N^+}} = {{O^+ + O^{+2}}\over{O^+}} \times {{He^+ +
He^{+2}}\over{He^+}},\\ 
ICF(Ne) &=& {{Ne}\over{Ne^{+2}}} = {{O^+ +
O^{+2}}\over{O^{+2}}} \times {{He^+ + He^{+2}}\over{He^+}},\\
ICF(S) &=& {{S}\over{S^+}+S^{+2}} = dexp[-.017+(.18 \times \beta)
-(.11 \times \beta^2) +(.072 \times \beta^3)],\\ 
ICF(Cl) &=& {{Cl}\over{Cl^{+2}+Cl^{+3}}} = {{He^+ + He^{+2}}\over{He^+}},\\
ICF(Ar) & = &{{Ar}\over{Ar^{+2}+Ar^{+3}}} = \frac{1}{1-\frac{N^+}{N}} \times {{He^+ + He^{+2}}\over{He^+}},
\end{eqnarray}
\end{mathletters}
where $\beta=log(O/O^{+})$ in eq.~2e. Also, when $\log (O/O^+)\le
+0.6$, our model tests suggested that Ar/Ar$^{+2}$=1, and so we
used this relation to obtain elemental argon abundances under that
condition.

Finally, we tested our abundance methods by comparing model input abundances with
derived abundances for S, Cl, and Ar, the elements we are focusing on
in this paper. To accomplish this, we used ABUN, including the ICFs listed
above, to calculate abundances based on the output line strengths
in our grid. We then divided each derived
abundance into the input abundance set in the same model. These
quantities are then plotted logarithmically in Fig.~4 against central
star temperature for the radiation-bounded models calculated out to
the Str{\"o}mgren radius (solid lines) as well as the matter-bounded
ones calculated out to a distance of one-half the Str{\"o}mgren radius
(dashed lines). If our scheme were perfect, lines in all panels would
be exactly horizontal at a constant ordinate value of zero. With a few
exceptions, our methods consistently produce abundances for these
elements within 0.1~dex of the actual abundance in the model, strongly
suggesting that the methods we apply to the data for the
PNe in our sample below are reliable.

\section{Results}

Derived electron temperatures and densities for our objects are listed in the last six rows of
Tables~5A-D. Generally speaking, these values are uncertain by
$\pm$10\%. Note that in most instances, [O~III], [N~II], and [S~III]
temperatures agree closely, indicating that our measurements of the
weak auroral lines in those cases were quite good. A few improbable
values for temperature and density are enclosed in parentheses; likely
causes are measurement uncertainties in the lines used and/or the effect of
errors in the reddening correction over a long wavelength baseline.

Abundances for observed ions are reported in Tables~5A-D, along with
ICFs for each element. The first column of each table contains the ion
ratio or ICF whose value is given in the subsequent columns for the
objects shown at the column head. We point out that these are {\it
number} abundance ratios. Also, note that while we observed [Ar~V]
$\lambda$7005 in most of our objects, its strength is generally very
weak, and photoionization models indicate that in PNe Ar$^{+4}$ is not
expected to be present at sufficient levels to influence the Ar
abundance measurement. Hence, we do not report an abundance for this ion.

In the case of S$^{+2}$/H$^+$, we provide two values. The first one is
the value derived using the NIR lines of [S~III]
$\lambda\lambda$9069,9532 (as described in {\S}3), while the second
one is derived using the [S~III] $\lambda$6312 line\footnote{Note that
the [N~II] temperature was employed when S$^{+2}$ was derived from the
$\lambda$6312 line, whereas the [S~III] temperature was used in association with the NIR lines, as mentioned in {\S }3.1}.  We note that these two sets of numbers agree
well in most cases. Fig.~5 is a plot of S$^{+2}$/H$^+$ derived from
the $\lambda$6312 line versus the same ratio derived from the NIR
lines. The diagonal line shows the track for a one-to-one
correspondence. Clearly the agreement is good in all but two cases,
M1-54 and M1-57, where $\lambda$6312 results for $S^{+2}/H^+ \times
10^6$ exceed 10 in Fig.~5.  These two discrepancies can be traced to
severe blending problems of [O~I] $\lambda$6300 with [S~III]
$\lambda$6312; in these two objects the ratio of $\lambda$6300 to
$\lambda$6312 is extremely high.  Generally, however, the agreement in
S$^{+2}$ abundances derived from $\lambda$6312 and the NIR lines
indicates that the common use of the [S~III] $\lambda$6312 line in the
past for computing sulfur abundances, e.g., Kingsburgh \& Barlow
(1994), is validated.

Finally, note that in many cases the ICF for nitrogen is very
high. This can be directly traced to the relatively large value for
the ratio of O/O$^+$ which is a factor in ICF(N) (see eq.~2C).  This
is likely due to the fact that these PNe are truncated or matter
bounded, causing the O/O$^+$ ratio to be large, and, likewise N/N$^+$. This may in turn impact the Ar abundances through ICF(Ar), causing the former to be underestimated in these cases.

Elemental abundances with respect to H$^+$ and O are reported in
Tables~6A-D.  Note that the last two columns of Tables~6 contain
abundances measured in the Sun by Grevesse et al. (1996) and in the
Orion Nebula by Esteban et al. (1998) for purposes of comparison.

Many of our objects are included in the abundance compilation
published by Maciel \& K{\"o}ppen (1994), and comparing our observed line
intensities with those from sources listed in that paper reveals
good agreement in general. Variations in ionic abundances provided in those same sources can at least
partially be be accounted for by differences in the adopted values of
the reddening constant, {\it c}, and in the extinction functions
employed. For example, at moderate reddening, given equal
$\lambda$3727/H$\beta$ flux ratios, a difference of 0.25 in the value of
{\it c} and of 0.025 in the value of f($\lambda$) can produce a
difference of $\sim$20\% in the corrected intensity ratio and thus in
O$^+$/H$^+$ abundance. Similarly, for equal $\lambda$9532/H$\beta$
ratios, a difference of 0.25 in {\it c} and of 0.08 in f($\lambda$)
can yield a corrected intensity ratio and inferred S$^{+2}$/H$^+$
abundance different by $\sim$65\%.

Our final abundances as listed in Table~6A-D agree well with previous
determinations, almost exclusively within the quoted errors. For
example, our values for NGC~6439 and IC~5217 are quite close to those
determined by Freitas-Pacheco, Maciel \& Costa (1992), and by Hyung et
al. (2001), respectively. Disagreements in other cases may be due to
analytical differences, as described above, as well as to
limitations in the spectral resolution and range of previous
observations.

While we are postponing a detailed analysis of our derived abundances
until Paper~V of this series when the results of the entire sample are
available, we present abundance results for S/O, Cl/O, and Ar/O for
our current sample in Fig.~6.\footnote{Problems associated with the
well-documented discrepancy between abundances derived from forbidden
and from permitted lines (see, e.g., Peimbert, Luridiana, \&
Torres-Peimbert 1995; Mathis, Torres-Peimbert, \& Peimbert 1998;
Stasi{\'n}ska 1998; Garnett \& Dinerstein 2001) are circumvented to a
large extent by normalizing S, Cl, and Ar abundances to O, since then
we are comparing abundances derived from forbidden lines. Thus, while
abundances with respect to H$^+$ may be affected by this problem,
ratios of other metals with respect to oxygen should remain
unaffected.} The top panel is a plot of S/O versus O/H for our
objects.  We also show the position of the Sun (Grevesse et al. 1996;
star) and the Orion Nebula (Esteban et al. 1998; X) for comparison.
The middle and bottom panels share the same format as the top one but
for the ordinates, which are, respectively, Cl/O, and Ar/O. A vertical
error bar in each panel indicates the ordinate uncertainty, while the
horizontal error bar in the top panel serves to show the uncertainty
in O/H for all three panels. Table~7 compares arithmetic averages and
standard deviations of the abundance ratios in our sample with
averages for PN samples reported by Kingsburgh \& Barlow (1994) and
Aller \& Keyes (1987), as well as for the Sun and Orion Nebula.

For S, we see that our average S/O is slightly lower than the findings
for the other PN samples as well as for Orion and the Sun. However,
our results are consistent with those of Kingsburgh \& Barlow,
considering the uncertainties associated with both samples. [We also note that Stasi{\'n}ska's expression ($\alpha =
3$) is the one adopted by Kingsburgh \& Barlow (1994).] We
speculate that systematically lower values for S/O in our sample are at
least partially related to our use of an ICF(S) relation (see eq. 2e)
that tends to yield corrections that are smaller than those used in
the comparison samples and for Orion (see discussion of Fig.~4 above). We mention again that this is perhaps related to the inclusion of dielectronic recombination in the derivation of ICF(S) for the first time. Once accurate cross-sections for this process are available, the behavior of ICF(S) may change significantly, and this fact must be kept in mind when considering our final sulfur abundances presented here.

Final values of Cl/O and Ar/O show
greater scatter and larger uncertainty, and this result is undoubtedly
related to the fact that generally the line strengths used to
determine Cl and Ar abundances are much weaker than those for
S. However, our averages for Cl and Ar compare favorably with those
listed in Table~7 from the other sources.

\section{Summary}

We present the first large spectroscopic survey of Type II planetary
nebulae in which the spectral range extends from 3600-9600{\AA}. This
wide range includes the strong nebular lines of [S~III]
$\lambda\lambda$9069,9532, allowing us to measure ionic abundances of
S$^{+2}$, and by extension, accurate abundances of elemental S. In
addition, most spectra contain lines of [Cl~III] $\lambda$5537,
[Cl~IV] $\lambda$8045, [Ar~III] $\lambda$7135, and [Ar~IV]
$\lambda$4740 which allow us to determine elemental abundances of Cl
and Ar as well.

The method we used to determine abundances includes a 5-level atom
program which determines ionic abundances of observed ions, then
applies an ionization correction factor to the sum of these ions to
render an elemental abundance. The principal challenge of this
research has been to establish a reliable, robust ionization
correction factor for S. Using a grid of photoionization models
spanning a broad range in stellar effective temperature, nebular
density, and degree of matter-boundedness, we established a relation
between the ratio of O/O$^+$ and S/(S$^+$+S$^{+2}$) that could be
confidently employed in our conversion of ionic abundances to
elemental abundances for S. In the same fashion we determined
correction factors for Cl and Ar.

In this paper we present our results for 18 PNe, all located in the
northern hemisphere. Our abundance results are consistent with general
trends seen before in S/O, Cl/O, and Ar/O. We observe slightly lower
S/O ratios than do some earlier surveys of PNe and H~II regions,
although the significance of this difference is questionable because
of overall uncertainties in measured abundances. We show that S$^{+2}$
abundances determined using the NIR lines of [S~III] are quite
consistent with those inferred from the analysis of the [S~III]
$\lambda$6312 line. Thus, sulfur abundances in the literature which
were derived from only the $\lambda$6312 line appear to be reliable
when this line is well measured. This agreement also implies that
simple steps described here to minimize atmospheric absorption effects
on the results from the NIR [S~III] lines are generally sufficient to
allow confident use of these lines.

\acknowledgments

We thank the referee, Walter Maciel, for his comments, particularly
regarding the sulfur ICF; we also thank Harriet Dinerstein for reading
and commenting on the manuscript. We are grateful to the staff at KPNO
for granting us observing time and to the IRAF staff for their ready
answers. We thank Joel Iams (Williams '01) and Hugh Crowl (Wesleyan
'00) for their help in obtaining the data. We also thank Marshall
McCall for a useful correspondence pertaining to the atmospheric
problems encountered in the near IR spectral region. KBK acknowledges
computer support by John Markunas of the Williams College OIT. This
research is supported by NSF grant AST-9819123.

\clearpage



\clearpage

\begin{figure}
\figurenum{1}
\plotone{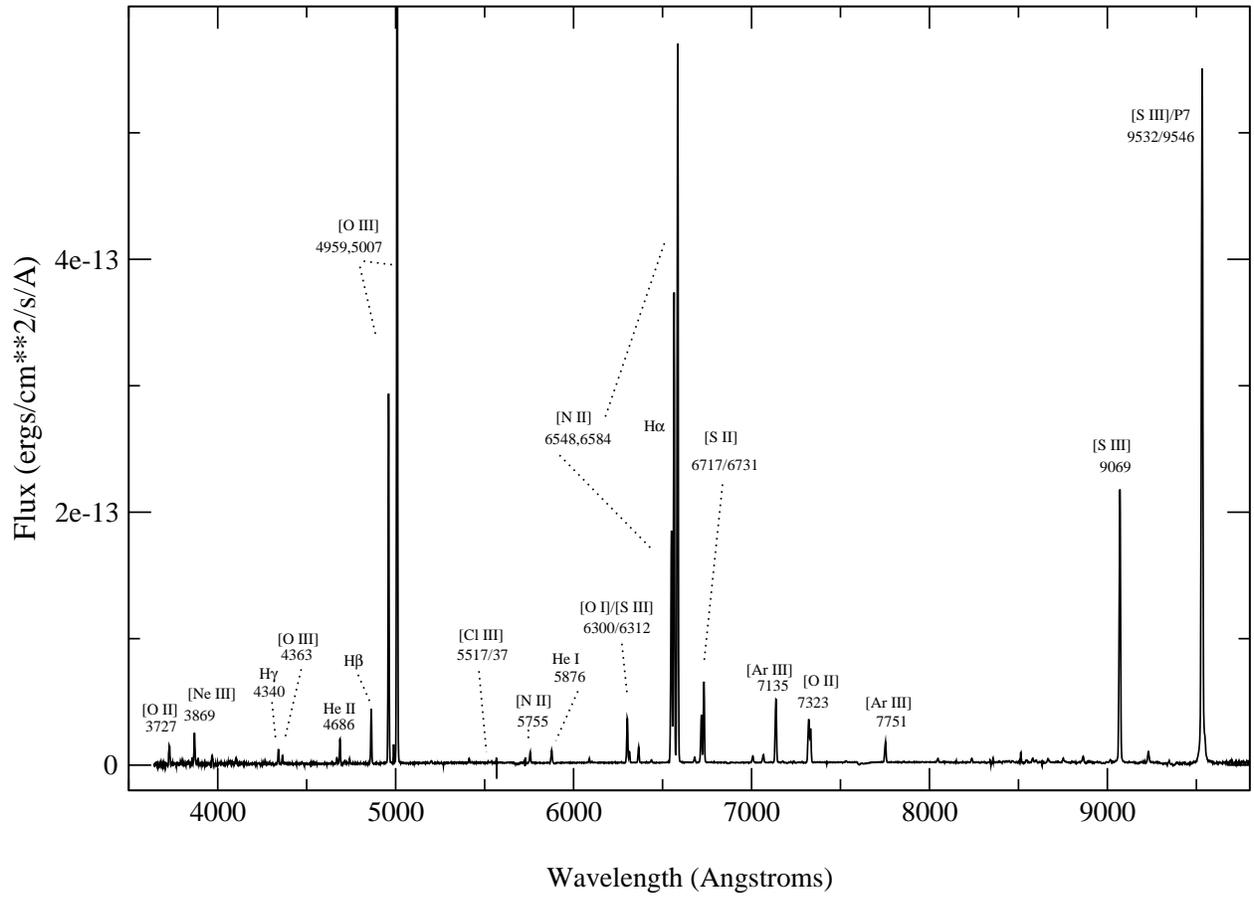}
\caption{Spectrum of M1-57, produced by merging blue and red spectra observed for the object. Important emission lines used to determine diagnostics and abundances are indicated.}
\end{figure}

\begin{figure}
\figurenum{2}
\plotone{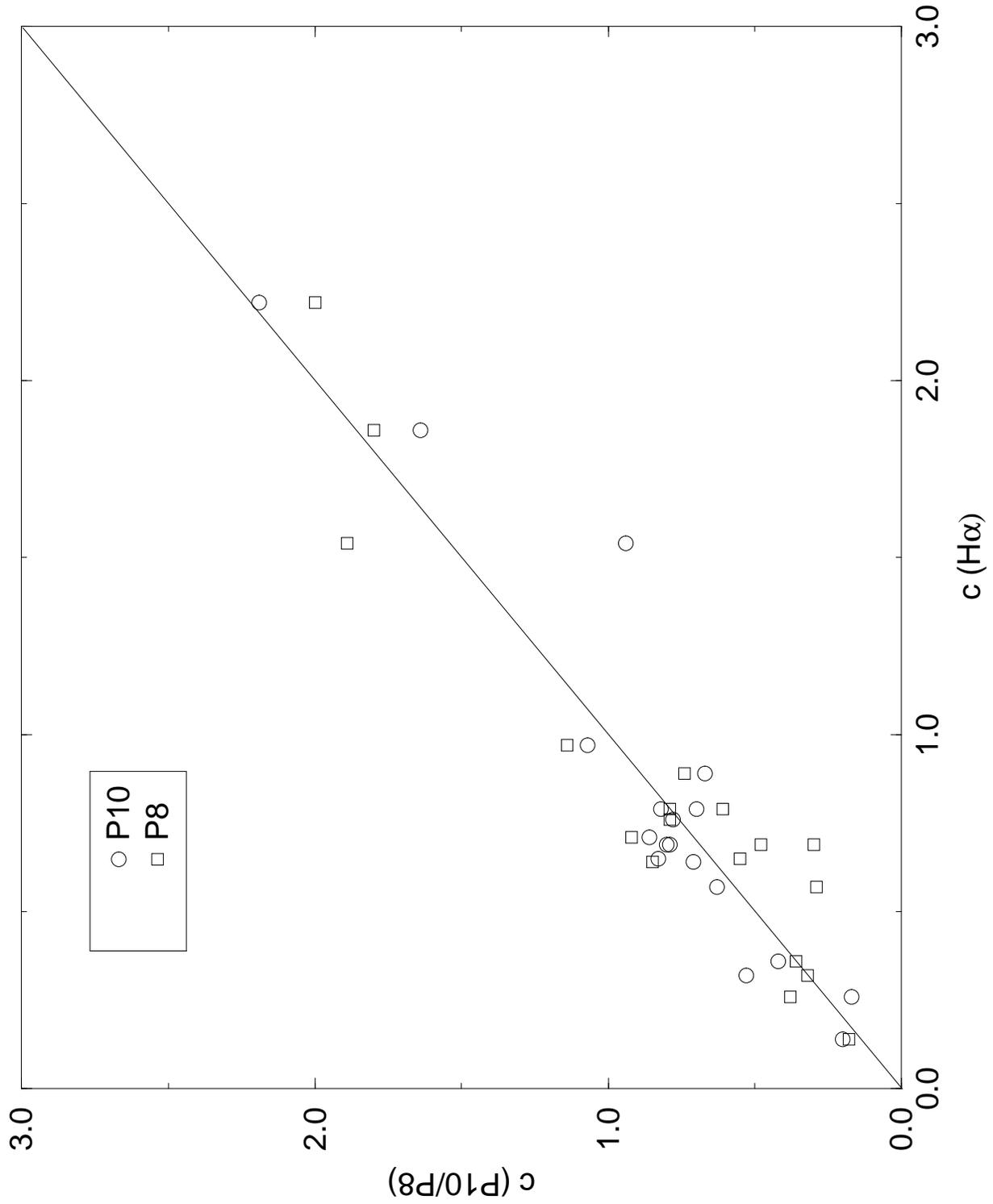}
\caption{A comparison of logarithmic extinction {\it c} as determined
using the Paschen 8 and 10 lines versus the value inferred from using
H$\alpha$. The solid line shows the track for a one-to-one
correspondence.}
\end{figure}

\begin{figure}
\figurenum{3}
\plotone{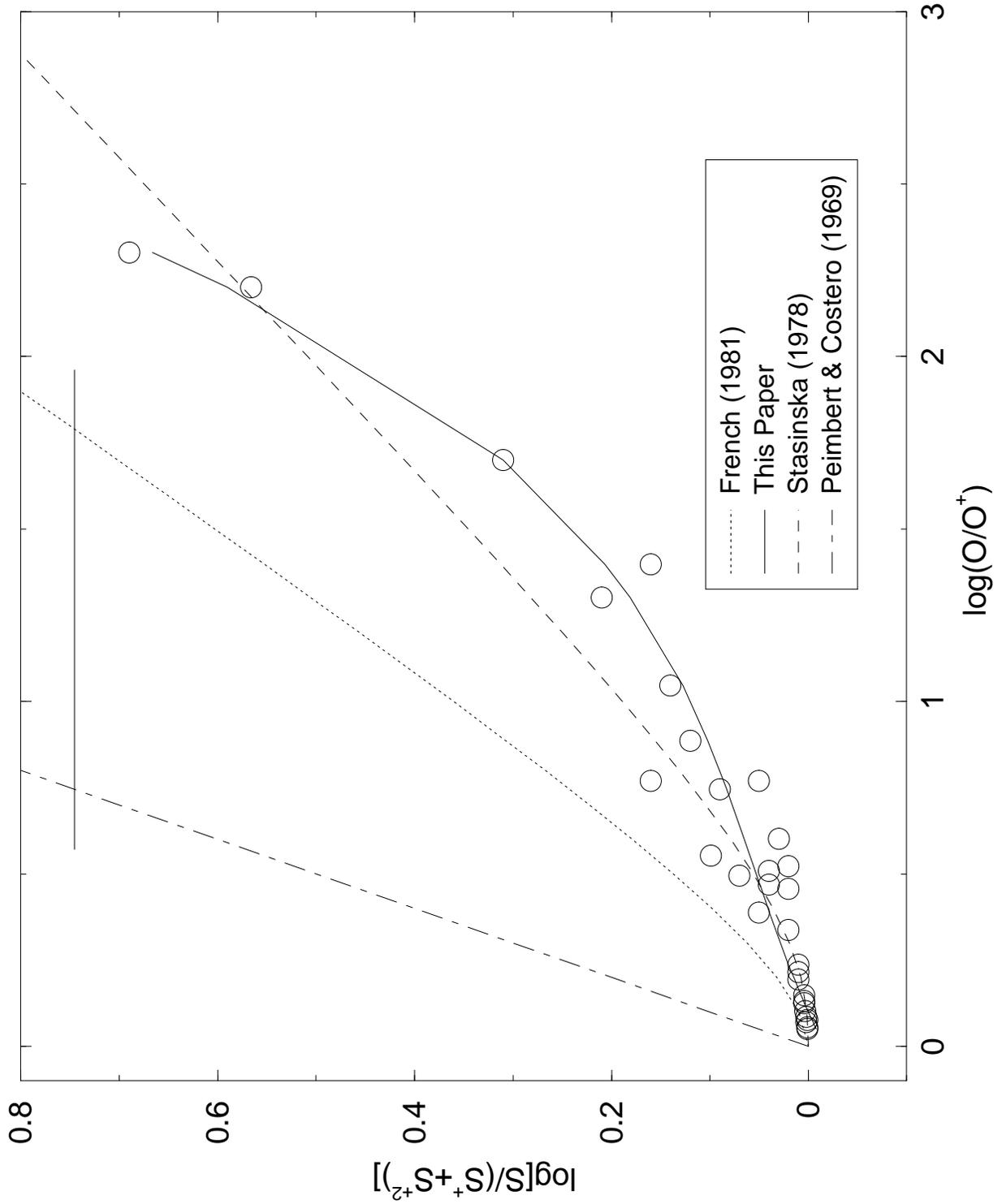}
\caption{Logarithmic ratios of the elemental S abundance relative to the
sum of S$^+$ and S$^{+2}$ ionic abundances vs. elemental O to the
O$^+$ ionic abundance from predictions of photoionization models
spanning a range in stellar effective temperature, nebular density,
and matter-boundedness, as discussed in the text. Each circle
represents one model. The solid horizontal line at the top of the
graph shows the range of our observed values of O/O$^+$.  The solid
curve is our third order fit whose coefficients are given in the text,
while the dot-dashed, dotted and dashed lines correspond to relations
developed by Peimbert \& Costero (1969), French (1981) and
Stasi{\'n}ska (1978), respectively.}
\end{figure}

\begin{figure}
\figurenum{4}
\plotone{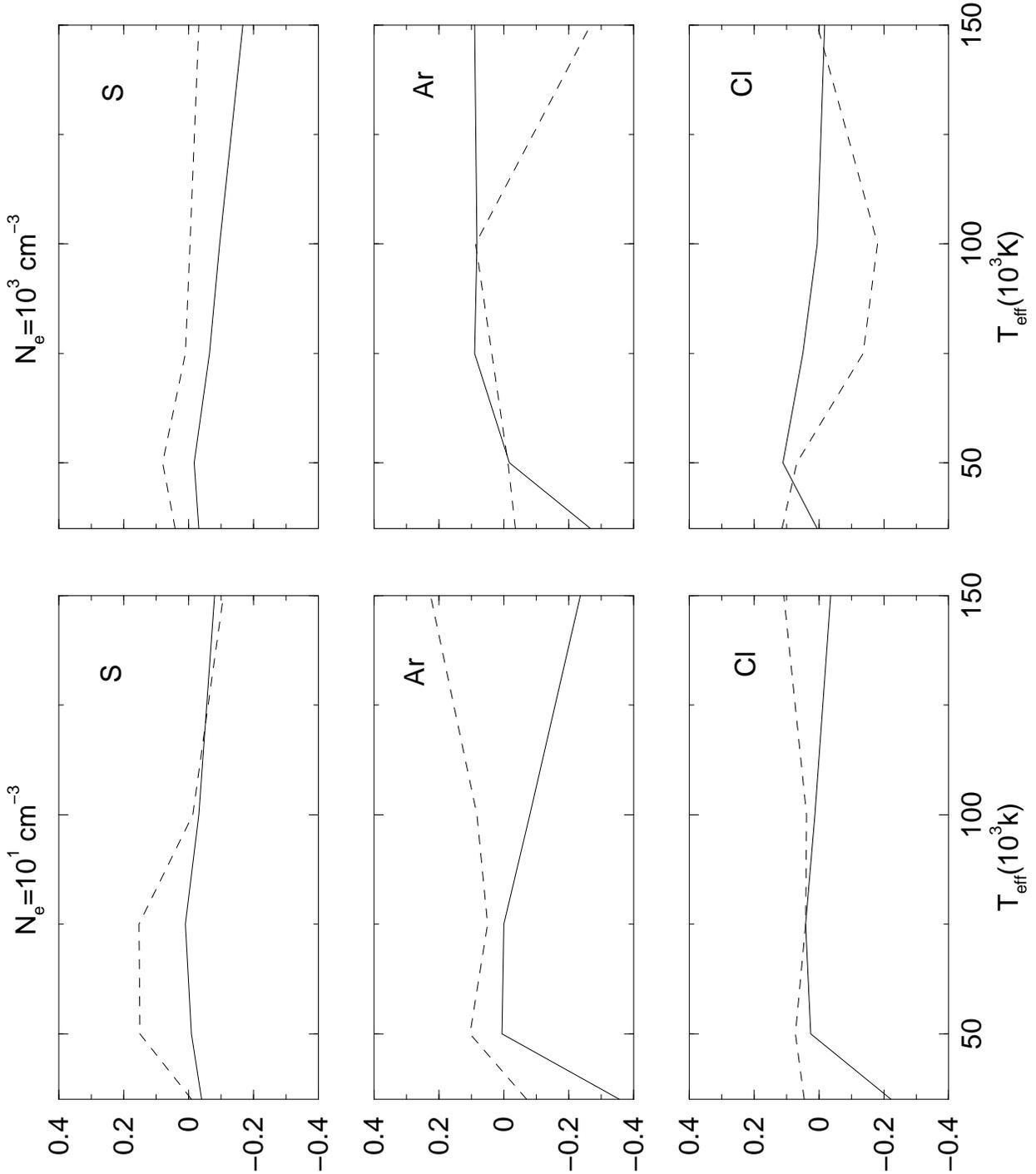}
\caption{Logarithmic ratio of model elemental abundance inferred by
ABUN relative to the actual model input abundance for S, Ar, and Cl,
vs. model stellar effective temperature. Plots on the left and
right are for nebular densities of 10 and 1000 cm$^{-3}$,
respectively. Solid curves show track of models which extend out to 1
Str\"{o}mgren radius, while the dashed lines show the same for models
which are stopped at 0.5 Str\"{o}mgren radius.}
\end{figure}

\begin{figure}
\figurenum{5} \plotone{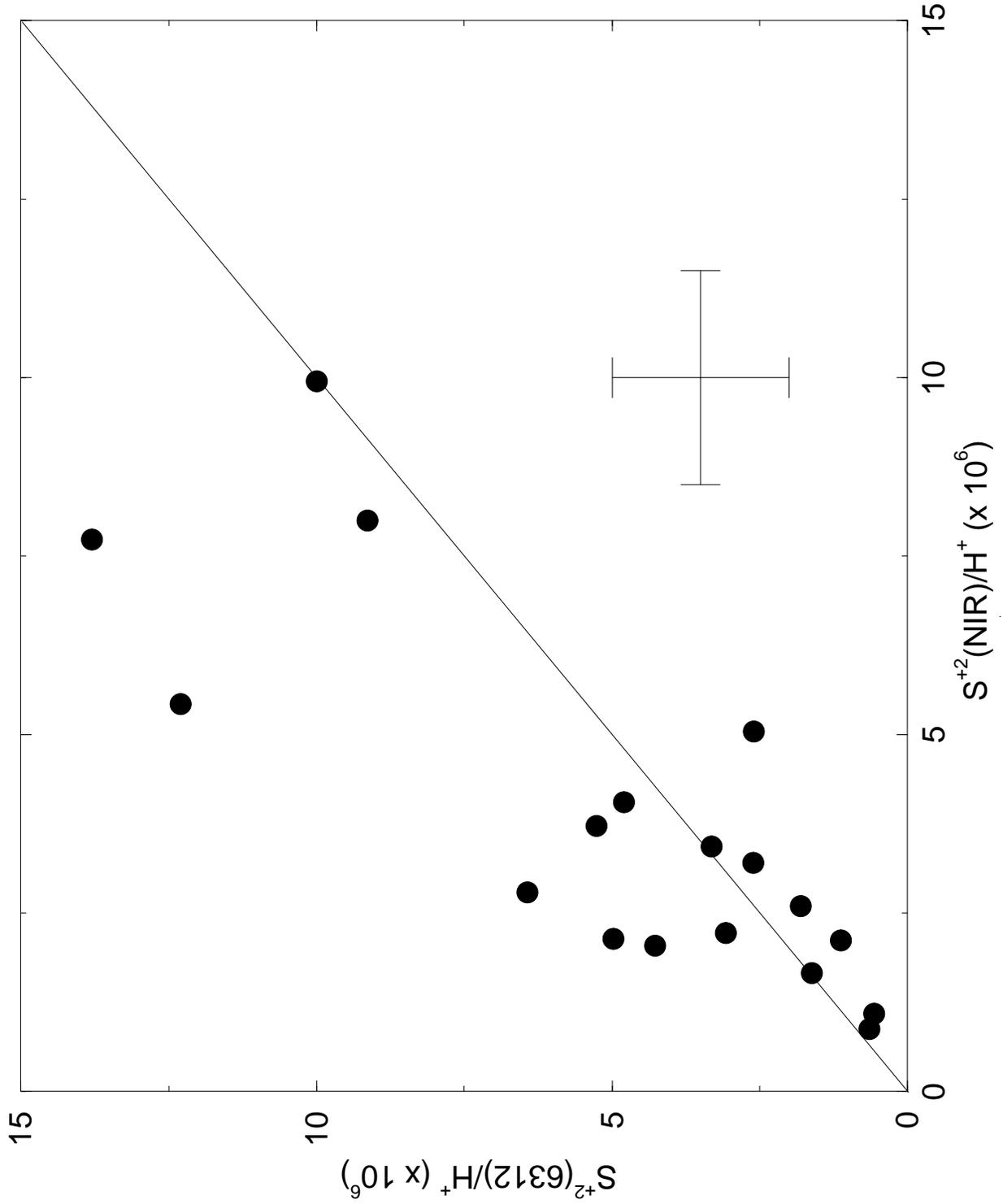}
\caption{Comparison of S$^{+2}$/H$^+$ for S abundances computed using
the 6312{\AA} emission line along with the [N~II] temperature
(ordinate) and the NIR emission lines along with the [S~III]
temperature (absissa). The solid line shows the track of a one-to-one
correspondence.}
\end{figure}

\begin{figure}
\figurenum{6}
\plotone{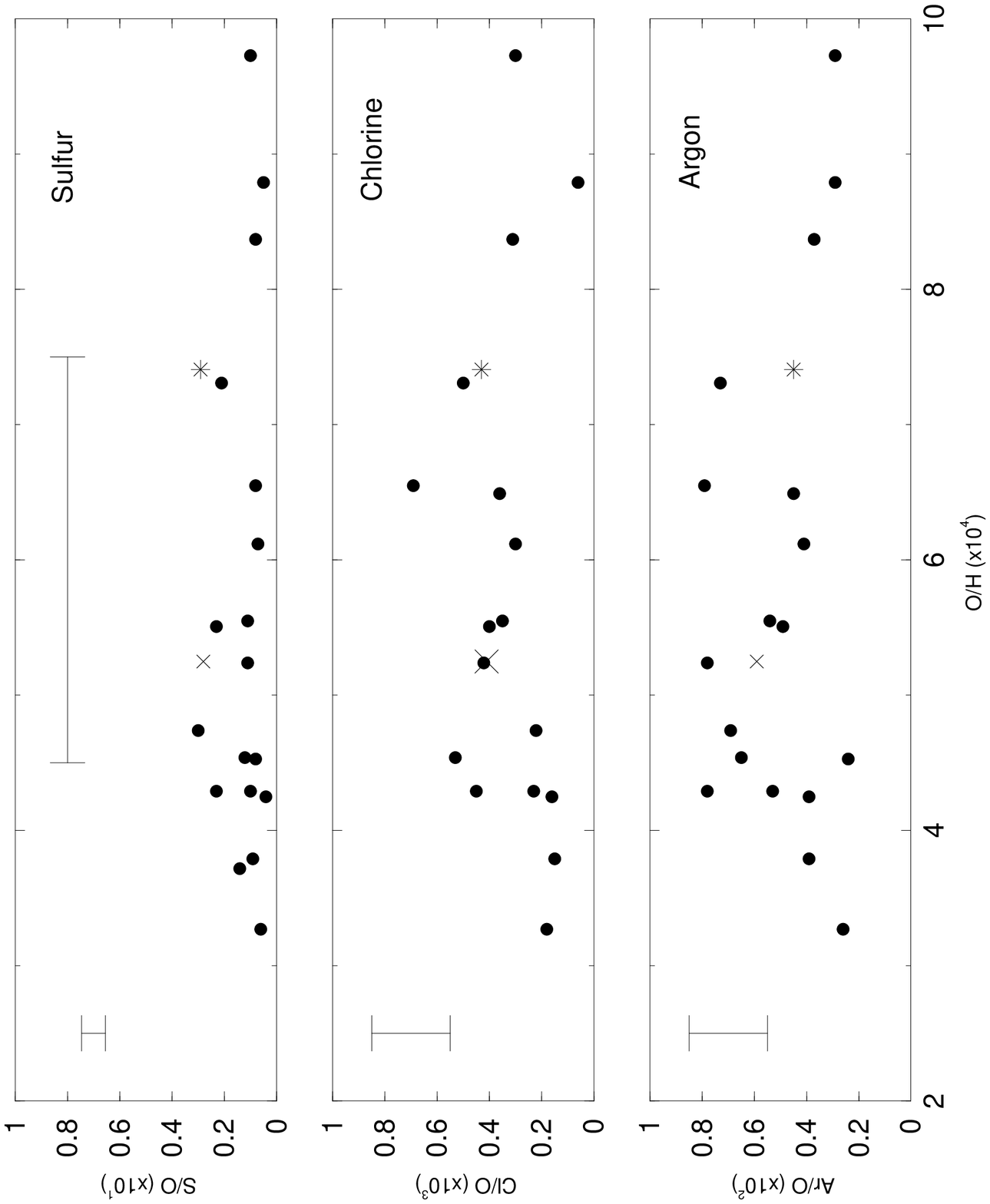}
\caption{Top: S/O (x 10$^1$) versus O/H (x 10$^4$), where filled
circles are ratios determined in this paper. The position of the sun
(Grevesse et al. 1996) and the Orion nebula (Esteban et al. 1998) are
indicated with a star and an X, respectively. Middle: Same as top but
for Cl/O (x 10$^3$). Bottom: Same as top but for Ar/O (x
10$^2$). Ordinate uncertainties are shown with error bars in each
panel, while the horizontal error bar in the top panel shows the O/H
uncertainty for all three panels.}
\end{figure}


\clearpage

\begin{thebibliography}{}  

\bibitem[]{} Acker, A., Ochsenbein, F., Stenholm, B., Tylenda, R.,
Marcout, J., \& Schohn, C. 1992, The Strasbourg-ESO Catalogue of
Galactic Planetary Nebulae (Garching: ESO)
\bibitem[]{} Aldrovandi, S.M.V., \& P{\'e}quignot, D. 1973, \aap, 25, 137
\bibitem[]{} Ali, B., Blum, R.D., Bumgardner, T.E., Cranmer, S.R., Ferland, G.J., Haefner, R.I., \& Tiede, G.P. 1991, \pasp, 103, 1182
\bibitem[]{} Aller, L.H., \& Czyzak, S. 1983, \apjs, 51, 211
\bibitem[]{} Aller, L.H., \& Keyes, C.D. 1987, \apjs, 65, 405
\bibitem[]{} Baluja, K.L., \& Zeippen, C.J. 1988, J. Phys. B, 21, 1455
\bibitem[]{} Barker, T., 1978a, \apj, 220, 193
\bibitem[]{} ------. 1978b, \apj, 221, 145
\bibitem[]{} ------. 1980a, \apj, 237, 482
\bibitem[]{} ------. 1980b, \apj, 240, 99
\bibitem[]{} ------. 1983, \apj, 270, 641
\bibitem[]{} Bhatia, A. K., \& Kastner, S. O. 1995, \mnras, 272, 311
\bibitem[]{} Burke, V.M., Lennon, D.J., \& Seaton, M.J. 1989, \mnras,
236, 353
\bibitem[]{} Butler, S.E., \& Dalgarno, A. 1980, \apj, 241, 838
\bibitem[]{} Butler, K., \& Zieppen, C.J. 1989, \aap, 208, 337
\bibitem[]{} ------. 1994, \aaps, 107, 1
\bibitem[]{} Cappellaro, E. Evans, R., \& Turatto, M. 1999, \aap, 351, 459
\bibitem[]{} Chapman, R.D., \& Henry, R.J.W. 1971, \apj, 168, 169
\bibitem[]{} Clegg, R. 1987, \mnras, 229, 31p
\bibitem[]{} Dennefeld, M., \&  Stasi{\'n}ska, G. 1983, \aap, 118, 234
\bibitem[]{} Dinerstein, H., 1980, \apj, 237, 486
\bibitem[]{} Esteban, C., Peimbert, M., Torres-Peimbert, S., \& Escalante, V. 1998, \mnras, 295, 401
\bibitem[]{} Ferland, G.J. 1996, {\it Hazy, A Brief Introduction to Cloudy}, University of Kentucky Department of Physics \& Astronomy Internal Report
\bibitem[]{} Freitas-Pacheco, J.A 1993, \apj, 403, 673
\bibitem[]{} Freitas-Pacheco, J.A., Maciel, W.J., \& Costa, R.D.D. 1992, \aap, 261, 579
\bibitem[]{} French, H.B. 1981, \apj, 246, 434
\bibitem[]{}  Galav{\'i}s, M.E., Mendoza, C., \& Zeippen, C.J. 1995,
\aaps, 111, 347
\bibitem[]{} Garnett, D.R. 1989, \apj, 345, 282
\bibitem[]{} Garnett, D.R., \& Dinerstein, H. 2001, \apj, in press (astro-ph/0105206)
\bibitem[]{} Grevesse, N., Noels, A., \& Sauval, A.J. 1996, in ASP Conf. Ser. 99
, Cosmic Abundances, ed. S.S. Holt \& G. Sonneborn (San Francisco: ASP), 117
\bibitem[]{} Henry, R.B.C., Kwitter, K.B., \& Bates, J.A. 2000, \apj, 531, 928
\bibitem[]{} Henry, R.B.C., Kwitter, K.B., \& Howard, J.W. 1996, \apj, 458, 215
\bibitem[]{} Henry, R.B.C., \& Worthey, G. 1999, \pasp, 111, 919
\bibitem[]{} van den Hoek, L.B., \& Groenewegen, M.A.T. 1997, {\aap}S, 123, 305
\bibitem[]{} Hyung, S., Aller, L.H., Feibelman, W.A., \& Lee,
W.-B. 2001, \apj, in press
\bibitem[]{} Kingsburgh, R.L., \& Barlow, M.J. 1994, \mnras, 271, 257
\bibitem[]{} K{\"o}ppen, J.A., Acker, A., \& Stenholm, B. 1991, \aap,
248, 197
\bibitem[]{} Kwitter, K.B., \& Henry, R.B.C. 1996, \apj, 473, 304
\bibitem[]{} ------. 1998, \apj, 493, 247
\bibitem[]{} Lennon, D.J., \& Burke, V.M. 1994, \aaps, 103, 273
\bibitem[]{} Maciel, W.J., \& Chiappini, C. 1994, \apss, 219, 231
\bibitem[]{} Maciel, W.J., \&  K{\"o}ppen, J. 1994, \aap, 282, 436
\bibitem[]{} Maciel, W.J., \& Quireza, C. 1999, \aap, 345, 629
\bibitem[]{} Maeder, A. 1992, \aap, 264, 105
\bibitem[]{} Marigo, P., Bressan, A., \& Chiosi, C. 1998, \aap, 331,
564
\bibitem[]{} Mathis, J.S., Torres-Peimbert, S., \& Peimbert, M. 1998, \apj, 495, 328
\bibitem[]{} McLaughlin, B.M., \& Bell, K. L. 1993, \apj, 408, 753
\bibitem[]{} Mendoza, C. 1983, in IAU Symp. 103, {\it Planetary
Nebulae}, ed. D.R. Flower, Dordrecht: Reidel, 143.
\bibitem[]{} Mendoza, C., \& Zeippen, C.J. 1983, \mnras, 202, 981
\bibitem[]{} ------. 1982, \mnras, 198, 127
\bibitem[]{} Natta, A., Panagia, N., \& Preite-Martinez, A. 1980, \apj, 242, 596
\bibitem[]{} Nomoto, K., Hashimoto, M., Tsujimoto, T., Thielemann, F.-K., Kishimoto, N., Kubo, Y., \& Nakasato, N. 1997a, Nuc. Phys. A, 616, 79c
\bibitem[]{} Nomoto, K., Iwamoto, K., Nakasato, N., Thielemann, F.-K.,
Brachwitz, F., Tsujimoto, T., Kubo, Y., \& Kishimoto, N. 1997b,
Nuc. Phys. A, 621, 467c
\bibitem[]{} Osterbrock, D.E. 1989, {\it Astrophysics of Gaseous Nebulae and Active Galactic Nuclei}, (Mill Valley, CA: University Science Books)
\bibitem[]{} Pagel, B.E.J. 1978, \mnras, 183, 1p
\bibitem[]{} Peimbert, M. 1990, Rep. Prog. Phys., 53, 1559
\bibitem[]{} Peimbert, M., \& Costero, R. 1969, Bol. Obs. Tonantzintla y Tacubaya, 5, 3
\bibitem[]{} Peimbert, M., Luridiana, V., \& Torres-Peimbert, S. 1995,
Rev. Mex. Astron. y Astrof. Ser. Conf., Vol. 3, The Fifth Mexico-Texas
Conference on Astrophysics: Gaseous Nebulae and Star Formation,
p. 295
\bibitem[]{} Peimbert, M., \& Torres-Peimbert, S. 1987, Rev. Mex. Astron. Astrofis., 14, 540
\bibitem[]{} P{\'e}quignot, D.,  Petitjean, P., \&  Boisson, C. 1991, \aap, 251, 680
\bibitem[]{} Ramsbottom, C.A., Bell, K.L., \& Stafford, R.P. 1996,
Atom. Dat. Nuc. Dat. Tab. 63, 57
\bibitem[]{} Savage, B.D., \& Mathis, J.S. 1979, \araa, 17.73
\bibitem[]{} Seaton, M. J., Zeippen, C. J., Tully, J. A., Pradhan, A. K., Mendoza, C., Hibbert, A. 1992, Rev. Mex. Astr. Astrofis., 23, 19
 Berrington, K. A.
\bibitem[]{} Stasi{\'n}ska, G. 1978, \aap, 66, 257
\bibitem[]{} ------. 1982, \aaps, 48, 299
\bibitem[]{} ------. 1998, in ASP Conf. Ser. 147, Abundance Profiles: Diagnostic Tools for Galaxy History, ed. D. Friedli, M. Edmunds, C. Robert, \& L. Drissen (San Francisco: ASP), 142
\bibitem[]{} Stevenson, C.C. 1994, \mnras, 267, 904
\bibitem[]{} Storey, P. J., \& Hummer, D. G. 1995, \mnras, 272, 41
\bibitem[]{} Verner, D.A., Ferland, G.J., Korista, K.T., \& Yakovlev, D.G. 1996, \apj, 465, 487
\bibitem[]{} Wiese, W.L., Fuhr, J.R., \& Deters, T.M. 1996, in Atomic
Transition Probabilities of Carbon, Nitrogen, and Oxygen : A Critical
Data Compilation, Amer. Chem. Soc., Washington, D.C.
\bibitem[]{} Woosley, S.E., \& Weaver, T.A. 1995, \apjs, 101, 181
\bibitem[]{} Zeippen, C.J., Le Bourlot, J., \& Butler, K. 1987, \aaps,
188, 251

\end{thebibliography}
\end{document}